\documentclass[aps,prb,reprint,groupedaddress]{revtex4-1}
\usepackage{multirow}
\usepackage{graphicx}% Include figure files
\usepackage{subfigure}
\newcommand{\sro}{$\textrm{Sr}_{2}\textrm{RuO}_{4}$}
\newcommand{\sros}{$\textrm{Sr}_{2}\textrm{RuO}_{4}\;$}

\begin{document}

% You should use BibTeX and apsrev.bst for references
% Choosing a journal automatically selects the correct APS
% BibTeX style file (bst file), so only uncomment the line
% below if necessary.
%\bibliographystyle{apsrev4-1}

% You should use BibTeX and apsrev.bst for references
% Choosing a journal automatically selects the correct APS
% BibTeX style file (bst file), so only uncomment the line
% below if necessary.
%\bibliographystyle{apsrev4-1}

% Use the \preprint command to place your local institutional report
% number in the upper righthand corner of the title page in preprint mode.
% Multiple \preprint commands are allowed.
% Use the 'preprintnumbers' class option to override journal defaults
% to display numbers if necessary
%\preprint{}

\title{Perturbation theory analysis of the strain-dependent superconducting phase diagram for Sr$_2$RuO$_4$}

\author{J.J. Deisz}
%\email{jdeisz@callutheran.edu}
\affiliation{Department of Physics, California Lutheran University,
Thousand Oaks, California 91360, USA}
%\altaffiliation{permanent address}
% \affiliation{xxx University of Wuerzburg xxx}
%\thanks{}

\date{\today}

\begin{abstract}
Previously, it was shown that a superconducting state dominated by singlet $d_{x^2-y^2}$ intra-band pairs emerges from the fluctuation exchange approximation (FLEX) applied to a realistic model for \sro, a result that is increasingly alignment with experimental data. 
Here we apply FLEX to model the strain-dependent phase diagram of \sros and show that we are able to reproduce its unusual features. This adds weight to the argument that a predominantly $d_{x^2-y^2}$ singlet pairing state represents a reasonable starting point for describing the superconducting properties of \sro.
\end{abstract}

% Make the title.
\maketitle 

\section{Introduction}

Thirty years since the discovery of superconductivity in strontium ruthenate \cite{maeno1994superconductivity}, \sro, a microscopic theory that convincingly explains a range of experimental results for this material remains elusive . The possibility of spin-triplet pairing was immediately recognized on
account of the observed enhancement ferromagnetic correlations in the normal state \cite{TMRice_1995} and the apparent absence of Knight shifts in NMR as temperature passes through the superconducting transition temperature \cite{ishida1998spin}. Additionally, muon spin rotation \cite{luke1998time} and Kerr effect \cite{PhysRevLett.97.167002} experiments provided evidence of time-reversal symmetry breaking in the superconducting state. These, and other confirming results, are consistently explained with a triplet pairing state having a $p_x + i p_y$ orbital structure. The possibility of such a pairing state has been supported by 
several model calculations using realistic Hamiltonian for \sro and various approximate treatments of the electron-electron interactions \cite{PhysRevLett.105.136401,Wang_2013,PhysRevB.89.220510, PhysRevLett.122.027002}. 

However, other experimental data appears to be inconsistent with $p_x + i p_y$ spin-triplet pairing. For example, the specific heat for temperatures below $T_c$ is suggestive of the existence of line nodes for the superconducting gap function on the Fermi surface \cite{nishizaki2000changes}.  Unexpectedly, reexamination of the Knight shift demonstrated the suppression of electron polarizability below $T_c$, in line with expectations for a spin-singlet superconductor \cite{pustogow2019constraints}. Further, the application of strain along the [1,0,0] crystalline axis, $\varepsilon_{xx}$, fails to produce a split in the superconducting transition that would be expected for
$p_x + i p_y$ pairing \cite{li2021high}. Indeed, the dramatic superconducting phase diagram for \sros
under strain provides an excellent target to enable microscopic models to finally provide clarity on the underlying pairing state for \sro \cite{hicks2014strong,steppke2017strong}.

In this manuscript we describe model results for the strain dependent phase diagram for \sros obtained using a realistic microscopic Hamiltonian and correlations approximated using the fluctuation exchange approximation (FLEX). Generically, we find that FLEX is able to reproduce two key results of the 
strain-dependent phase diagram: (1) there is no splitting of the superconducting transition as they symmetry between the $x$ and $y$ crystal axes is lifted and (2) $T_c$ plummets rapidly just after the strain exceeds the critical value where a van Hove singularity passes through the Fermi level, $\varepsilon_{xx} = \varepsilon_{\mathrm{vH}}$. Additionally, when the coupling strength is manually adjusted so that the calculated unstrained $T_c$ approaches that for experiment, FLEX accurately reproduces the strain-induced peak structure in $T_c$. Thus, we argue, the pairing symmetry generated by FLEX represents a strong candidate for describing the dominant pairing correlations in the superconducting state of \sro. 

\section{Model and numerical methods}

 We use a three atomic orbitals per unit cell basis, corresponding to the $4d_{xy}$, $4d_{xz}$ and $4d_{yz}$ orbitals of the ruthenium atoms, to account for the three distinct Fermi surface sheets (denoted  $\alpha$, $\beta$ and $\gamma$) observed experimentally \cite{PhysRevB.51.1385,Mackenzie1996510}.
Tight-binding hopping matrix elements, $t_{\nu,\nu^{\prime}}(\mathbf{R})$, where $\nu$ and $\nu^{\prime}$ are orbital indices, are taken from Pavarini and Mazin's \cite{PhysRevB.74.035115,PhysRevB.76.079901} fit to the density functional theory band-structure for the baseline unstrained case. An atomically local spin-orbit interaction, $\lambda \, \vec{s}\cdot \vec{l}$,
is assumed and we use the first-principles derived value of
$\lambda = 93\textrm{ meV}$ \cite{PhysRevLett.101.026406}. The chemical potential is adjusted to maintain an average filling of 2/3. Finally, the relatively small interplanar hopping terms are ignored consistent with the quasi two-dimensional behavior of \sro. 

Electron correlations are modeled starting with
an atomically-local electron-electron interaction vertex, $\Gamma^{(0),\textrm{cRPA}}$, 
parameterized with band-dependent intraorbital ($U_{\nu}$) and interorbital 
($U^{\prime}_{\nu\ne \nu^{\prime}}$) Coulomb 
and exchange ($J_{\nu\ne \nu^{\prime}}$) interaction terms evaluated for \sro~using the
constrained random phase approximation calculation \cite{PhysRevB.86.165105}. 
The largest interaction parameters, \textit{i.e.} $U_{xy} = 2.72\textrm{ eV}$ and $U_{xz} = 2.48\textrm{ eV}$,
are comparable in size to the unrenormalized bandwidth suggesting that \sro~is in the intermediate-coupling regime. 
Energy renormalization and lifetime broadening of quasiparticle excitations are calculated via the quasiparticle equation
\begin{widetext}
\begin{equation}
\sum_{\nu^{\prime}\sigma^{\prime}} \left(H^{(0)}_{\nu\sigma;\nu^{\prime}\sigma^{\prime}}(\mathbf{k}) + \Sigma_{\nu\sigma;\nu^{\prime}\sigma^{\prime}}(\mathbf{k},E_{qp}+i0^+)\right) 
\psi_{\nu^{\prime}\sigma^{\prime}}(\mathbf{k},E_{qp}) = (E_{qp}+i\Gamma_{qp})\,\psi_{\nu\sigma}(\mathbf{k},E_{qp})
\end{equation}
\end{widetext}
where
$\Sigma_{\nu\sigma;\nu^{\prime}\sigma^{\prime}}(\mathbf{k},E)$ is the electron self-energy and $\nu$ and $\sigma$ are orbital and spin indices respectively. 

The self-energy is approximated using the self-consistent Fluctuation Exchange Approximation (FLEX) develop by Bickers, White and Scalapino \cite{PhysRevB.43.8044} supplemented by the dynamic cluster approximation (DCA) \cite{PhysRevB.58.R7475} and generalized to explore the superconducting state below $T_c$ \cite{PhysRevB.80.094516}. We use a $256 \times 256$ unit cell/momentum space with a $4 \times 4$ DCA coarse-graining
of the self-energy to keep calculations computationally feasible while preserving the
momentum-dependence that is needed to describe 
correlation induced features that are unique for excitations near
the van Hove singularity \cite{PhysRevB.95.045122}.
As FLEX tends to overestimate the magnitude of the self-energy \cite{PhysRevB.50.403}, 
we introduce a single scale factor for the cRPA interaction vertex, \textit{i.e}
$\Gamma^{(0),FLEX} = g_o \, \Gamma^{(0),cRPA}$ with $g_o < 1$. 

Strain along the [1,0,0] direction drives a van Hove singularity in the $\gamma$-band from its unstrained energy of $E_{vH}\sim 20$~meV above the Fermi level, through and below the Fermi level. This
evolution of a high density-of-states feature in the bandstructure
is the likely source of the dramatic concurrent variation in $T_c$ \cite{hicks2014strong,steppke2017strong}.
We use a dimensionless parameter,
 $\eta$, to simply represent the strain-driven changes in the tight-binding parameters most relevant for this 
 process.
 We have
\begin{eqnarray}
\tilde{t}_{xy,xy}(\pm 1, 0) & = & 
 (1-\eta)\; t_{xy,xy}(\pm 1, 0) \\
\tilde{t}_{xy,xy}(0, \pm 1) & = & (1+\eta) \; t_{xy,xy}(0, \pm 1) 
\end{eqnarray}
where $\tilde{t}$ and $t$ are the strained and unstrained hopping values respectively. Effectively,
this is consistent with tensile strain along the [1,0,0] direction
As $\eta$ becomes sufficiently large, the van Hove singularity at $\mathbf{k}= (0,\pm \pi)$ passes through the Fermi level at some $\eta_{\mathrm{vH}}$ and will be pushed increasingly below the Fermi level for $\eta > \eta_{\mathrm{vH}}$. Assuming that tight-binding parameters vary linearly with strain, then our model parameter should track experiment via $|\eta|/\eta_{\mathrm{vH}} = |\varepsilon_{xx}|/\varepsilon_{\mathrm{vH}}$.

\section{Results without strain}

When FLEX is applied to this model
of \sros in the unstrained limit, the pairing symmetry that results \cite{PhysRevLett.107.277003} \textit{is not} chiral, spin-triplet $p$-wave. 
In the absence of spin-orbit coupling ($\lambda=0$) FLEX generates a superconducting state consisting of singlet $d_{x^2-y^2}$ pairs dominated by 
intraband pairing of quasiparticles in the quasi-two-dimensional $\gamma$-band. When the spin-orbit interaction is included ($\lambda=93$~meV), the pairing state acquires triplet components representing interorbital pairing across all three bands. However, these triplet-pairing contributions remain small in comparison to the dominant singlet $d_{x^2-y^2}$ pairing terms. 

The quasiparticle excitations at and near the van Hove singularity in the $\gamma$-band clearly appear to be essential drivers of the strain dependence of $T_c$ for \sro~ and, perhaps, for superconductivity in the material more generally. A non-trivial representation of these quasiparticles is a strength of the FLEX method.  Indeed, FLEX results for these key $\gamma$-band excitations \cite{PhysRevB.95.045122}, obtained with a coupling strength of $g_o = 0.67$, are in excellent quantitative agreement with experimental results for the temperature and frequency dependencies of quasiparticle lifetimes \cite{PhysRevLett.94.107003}. Further, FLEX generates a downward shift of the van Hove energy from its DFT value of 90 meV to approximately 20 meV at $T=100$~K, in good agreement with the experimental result of 14 meV \cite{PhysRevLett.99.187001}. Although FLEX results for quasiparticle excitations demonstrate Fermi-liquid like behavior for \textit{nearly all} of the Fermi surface, there are hints of non-Ferm liquid behavior in the $\gamma$-band near the van Hove singularity, a possibility that would be consistent with recent experimental results for the Seebeck coefficient \cite{PhysRevB.105.184507}.

\begin{figure}
%    \centering
\includegraphics{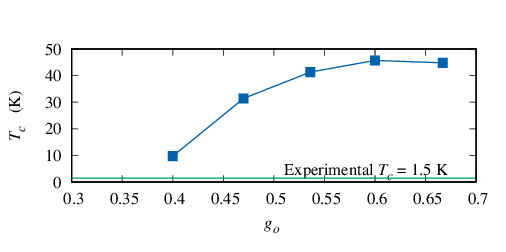}
    \caption{Calculated FLEX results for $T_c$ vs. dimensionless coupling strength parameter, $g_o$.  While experimental quasiparticle renormalizations are most accurately represented with $g_o \sim 0.67$, a $g_o$ value near 0.4 is more consistent with the observed $T_c$ for \sro.}
    \label{fig:Tc_vs_coupling}
\end{figure}

In Fig \ref{fig:Tc_vs_coupling} we show FLEX results for $T_c$ versus coupling strength $g_o$. The coupling strength most consistent with describing normal state quasiparticle excitatons, $g_o = 0.67$, produces a $T_c$ near 43~K, clearly much too high in comparison to the experimental value of approximately 1.5~K. This discrepancy is expected as some of the higher-order processes that are neglected in the FLEX approximation, such as fluctuations of a $d$-wave superconducting order parameter,
can be numerically insignificant for normal state quasiparticles, but essential
for calculating the superconducting $T_c$. 
The FLEX results for the unstrained $T_c$ is in alignment with
experiment when $g_o \simeq 0.4$. Indeed, we will
find that FLEX generates a strain-dependent phase diagram that accurately mimics experimental results 
when this lower coupling strength is used and strain
is varied.

\section{Results with strain}

\begin{figure}
    \centering
\includegraphics{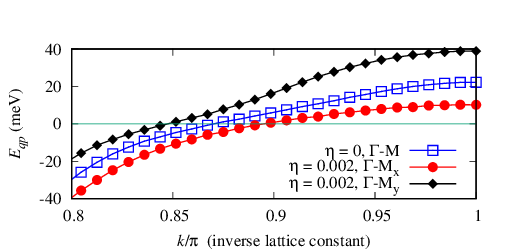}
    \caption{Quasiparticle energies, $E_{qp}$, vs. momentum $k/\pi$ along the cuts $\Gamma-M_{x}$ and $\Gamma-M_{y}$. In the unstrained case, $\eta = 0$, these cuts are degenerate, \textit{i.e.} $x=y$. In the strained case, $\eta = 0.002$, excitations along the $x$ ($y$) are lower (higher) in energy in comparison the unstrained case, with the van Hove singularity at $M_x$ driven toward the Fermi level
    where $E_{qp}=0$}
    \label{fig:dispersion}
\end{figure}

In the absence of strain, the van Hove singularity in the $\gamma$ band appears at symmetry equivalent $\mathbf{k}$-points $M_x = (\pm \pi,0)$ and $M_y = (0, \pm \pi)$. The dispersion near these points along the $\Gamma-M$ cut is displayed in Fig~\ref{fig:dispersion} for the unstrained, $\eta=0$ (open symbols), and strained cases, $\eta=0.002$ (closed symbols), at a coupling strength of $g_o = 0.67$. The energy at the van Hove singularity ($k/\pi = 1.0$) is 20~meV in the unstrained case. Finite strain splits the $\Gamma-M_x$ and $\Gamma-M_y$ dispersion curve with the $M_x$ van Hove energy lowered to 10~meV and the $M_y$ van Hove energy increased to 40~meV. The strain-lowered van Hove energy at $M_x$ is key as it passes through the Fermi level, $E_{F}=0$, with sufficient tensile strain along [1,0,0].

\begin{figure}
\centering
   \includegraphics{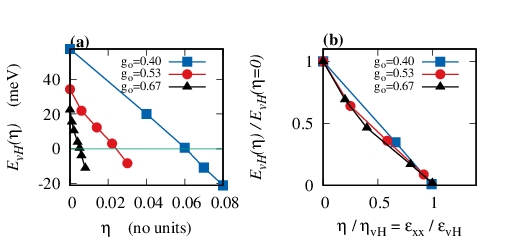}
\caption{
    (a) Quasiparticle energy at the van Hove singularity, $E_{\mathrm{vH}}$,  vs. strain parameter, $\eta$, at 
    various coupling strengths, $g_o$, and (b) $E_{\mathrm{vH}}$ normalized by its zero strain value vs. normalized strain parameter, $\eta/\eta_{\mathrm{vH}}=\varepsilon_{xx}/\varepsilon_{\mathrm{vH}}$. Stronger coupling leads to a 
    downward dynamical renormalization of $E_{\mathrm{vH}}(\eta)$ and, consequently, the critical strain value, $\eta_{\mathrm{vH}}$, where $E_{\mathrm{vH}} = 0$, is reduced as well. 
    However, $E_{\mathrm{vH}}$ vs. $\eta$ follows the same trends at all couplings as observed in the scaled plot.}
    \label{fig:quasi_vs_coupling}
\end{figure}

In Fig~\ref{fig:quasi_vs_coupling}(a) we show the $M_x$ quasiparticle energy as a function of strain, $\eta$, at different coupling strengths. For each coupling strength, a critical value of strain, $\eta_{\mathrm{vH}}$ is identified where the van Hove energy crosses the Fermi level. Since the $\gamma$-band is increasingly flattend and Hove energy is driven downward with an increase in the coupling strength, $\eta_{\mathrm{vH}}$ decreases as the coupling strength increases. 
Figure~\ref{fig:quasi_vs_coupling}(b) shows a scaled version of the same plot. Here $E_{\mathrm{vH}}(\eta)$ is divided by $E_{\mathrm{vH}}(\eta=0)$ and $\eta$ is expressed in terms of $\eta_{\mathrm{vH}}$; the plots then become similar for the coupling strengths considered here.

Superconducting properties of this model are described in FLEX via the equal-time anomolous Green's function:
\begin{equation}
<c_{\nu^{\prime}\sigma^{\prime}}(\mathbf{r}=0)
c_{\nu\sigma}(\mathbf{r})> \\
\equiv m_p(T)\,
\psi_{\nu\sigma;\nu^{\prime}\sigma^{\prime}}(\mathbf {r})
\end{equation}
Here $\psi$ is the normalized pair wave function and
$m_p(T)$ is the pair amplitude which becomes finite below $T_c$. As described in a previous work\cite{PhysRevLett.107.277003}, in the limit of zero strain the FLEX wave function for this model describes a superconducting state dominated by singlet, $d_{x^2-y^2}$ pairing of $\gamma$-band quasiparticles with minority triplet components induced by spin-orbit coupling, albeit in a manner that does not account for the breaking of time-reversal symmetry that is observed experimentally.

\begin{figure}
\centering
 \includegraphics{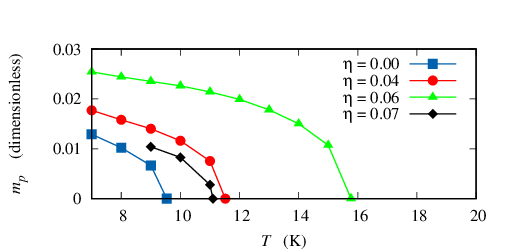}
\caption{
Superconducting order parameter, $m_p$, vs temperature, $T$, for several strain parameters at a 
coupling strength $g_o = 0.40$ for which $\eta_ {\mathrm{vH}}\sim 0.06$. The unstrained result, $\eta = 0.00$, 
shows a typical second order transition at $T_c \sim 9.5\textrm{K}$. 
For the intermediate strain value of $\eta = 0.04 < \eta_{\mathrm{vH}}$ an elevated $T_c \sim 11.5$~K is observed and smooth, 
single-phase behavior for $T< T_c$. 
At $\eta = 0.06 \sim \eta_{\mathrm{vH}}$, a maximum $T_c \sim 15.8$K is obtained and 
single phase behavior persists below the transition. The subsequent curve for $\eta = 0.07 > \eta_{\mathrm{vH}}$, 
shows a dramatic reduction in both $T_c$ and the overall strength of the order parameter.}
\label{fig:mp_vs_eta}
\end{figure}

In Fig~\ref{fig:mp_vs_eta} we show $m_p$ vs. $T$ curves for several different strain values at a coupling strength $g_o = 0.40$. The impact of strain on these curves is evident with the largest pair amplitudes and $T_c$ occuring at the critical strain value of $\eta_{\mathrm{vH}} = 0.06$. Superconductivity is quickly surpressed as $\eta$ increases beyond $\eta_{\mathrm{vH}}$.  

This method is able to explore the model for $T<T_c$ 
and, thus, address the possibility of an emergent second superconducting transition should a nearly-degenerate pairing symmetry become activated \cite{PhysRevB.75.064507}. Numerical evidence for a second transition consists of a discontinuous slope in the $m_p$ vs $T$ curves. These FLEX results, which appear to be analytic for $T<T_c$, are therefore consistent with a single-component superconducting state at all strain values. This result is consistent with heat capacity, elastocaloric effect and superfluid density measurements \cite{li2021high,li2022elastocaloric,mueller2023constraints}, but is inconsistent with $\mu$sr results that suggest a strain-induced splitting between superconducting states with and without time-reversal symmetry breaking \cite{grinenko2021split}. 

\begin{figure}
\begin{centering}
 \includegraphics{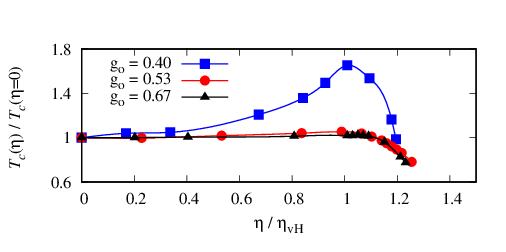}
\end{centering}
\caption{\label{Fig:Tc_vs_eta}
Relative superconducting transition temperature, $T_c(\eta)/T_c(\eta=0)$, versus normalized strain, $\eta/\eta_{\mathrm{vH}}$, for different coupling strengths, $g_o$. For all $g_o$, $T_c$ drops off rapidly for $\eta/\eta_{\mathrm{vH}} > 1$, consistent with what is observed for \sro. For $g_o = 0.40$, which corresponds to $T_c(\eta=0) \sim 9.5K$, we observe a pronounced peak in $T_c$ at or near $\eta/\eta_{\mathrm{vH}} \sim 1$ in alignment with experimental results for \sro. Thus, when the coupling strength is set to values consistent with the low $T_c$ values observed in \sro, FLEX-based three-band Hubbard model results accurately represent the unusual strain-dependent phase diagram for \sro. 
}
\end{figure}

The main result is shown in Fig~\ref{Fig:Tc_vs_eta} where $T_c$ as a function of strain at different coupling strengths is displayed. For all coupling strengths, the FLEX $T_c$ curves drop rapidly as strain increases beyond the critical value reproducing experimental results. The observed peak structure in $T_c$ vs. strain for \sros is also obtained with FLEX, but only for the lower coupling strength of $g_o = 0.40$. However, this coupling strength is most consistent with the low $T_c$ values observed for \sro. Apparently, the appearance of the peak in $T_c$ vs strain only emerges in these FLEX results when quasiparticle excitations at the van Hove singularity are sufficiently narrow in energy. 

Yuan, Bern and Kivelson showed that when using a BCS-like Hamiltonian with an assumed $d$-wave pairing interaction, mean-field theory also generates a peaked structure in $T_c$ vs. strain when  a similar model for the band-structure of \sro\cite{PhysRevB.108.014502}. 
Concurrent with the completion of this work, Hauck, \textit{et al.}, reported functional renormalization group results that suggest
of $d$-wave pairing correlations, as measured through calculation of $T=0$ pairing eigenvalues,
follow a similar trend \cite{hauck2023competition}.
Together with the results presented here, this suggests that the peaked structure in $T_c$ vs strain may be a generic feature of models with sharp quasiparticles in the $\gamma$-band forming $d$-wave singlet pairs. The results presented here are of added significance because (1) finite temperature and quasiparticle lifetime effects are included and (2) the pairing correlations emerge from a realistic Hamiltonian and we are able to track the $T< T_c$ physics.

\section{Discussion}

Starting from a realisitic microscopic Hamiltonian and an unbiased treatment of the electron correlations driving a superconducting transition, FLEX results account for (1) the apparent absence of a splitting of the superconducting transition in strained samples and (2) the peak structure observed in $T_c$ vs strain in \sro. This result adds to the body of evidence that the underlying pairing in \sro is dominated by quasiparticles in the $\gamma$-band bound in spin-singlet pairs having $d_{x^2-y^2}$ orbital symmetry. While FLEX results for the unstrained limit were published previously, experimental evidence has increasingly shifted to support this picture.

These FLEX results do not yet account for evidence of time-reversal symmetry breaking in unstrained samples below
$T_{tr} = T_c$ and in strained samples below $T_{tr} < T_c$. There are many potential explanations for this discrepancy. The additional pairing symmetry (or symmetries) needed to account for time-reversal symmetry breaking in a simple way may be quite small in comparison the the dominant $d_{x^2-y^2}$ term. 
If so, then the physical and numerical approximations of this method will make it difficult for these small terms to be resolved.
The momentum resolution we use in our numerical evaluation of the generalized self-energy (incorporating both normal and anomolous components) may be insufficient for accurately representing high angular momentum pairing components. Also,
fluctuations of the dominant $d_{x^2-y^2}$ order parameter are missing in FLEX, a significant shortcoming for a quasi two-dimensional system. 
Further, as is typically done for this system, the possibility of translational symmetry breaking in the superconducting state is not considered in our calculation scheme. 

A complete picture of superconductivity in \sros indeed does not emerge in these results, but they perhaps suggest a new path forward.
Empirically it may prove fruitful to treat the underlying superconductivity in \sros as emerging through a dominant single-component $d_{x^2-y^2}$ pairing state with time-reversal symmetry breaking being a secondary feature in this complex, multiband system.

\begin{acknowledgments}
We thank Fakher Assaad for useful conversations. This work was supported by the W\"urzburg-Dresden
  Cluster of Excellence on Complexity and Topology in Quantum Matter ct.qmat
  (EXC 2147, project-id 390858490) and SCIAS (Siebold-Collegium, Institute for Advanced Studies, University of Würzburg).

\end{acknowledgments}

% Create the reference section using BibTeX:
%\bibliography{new_refs}

\begin{thebibliography}{35}%
\makeatletter
\providecommand \@ifxundefined [1]{%
 \@ifx{#1\undefined}
}%
\providecommand \@ifnum [1]{%
 \ifnum #1\expandafter \@firstoftwo
 \else \expandafter \@secondoftwo
 \fi
}%
\providecommand \@ifx [1]{%
 \ifx #1\expandafter \@firstoftwo
 \else \expandafter \@secondoftwo
 \fi
}%
\providecommand \natexlab [1]{#1}%
\providecommand \enquote  [1]{``#1''}%
\providecommand \bibnamefont  [1]{#1}%
\providecommand \bibfnamefont [1]{#1}%
\providecommand \citenamefont [1]{#1}%
\providecommand \href@noop [0]{\@secondoftwo}%
\providecommand \href [0]{\begingroup \@sanitize@url \@href}%
\providecommand \@href[1]{\@@startlink{#1}\@@href}%
\providecommand \@@href[1]{\endgroup#1\@@endlink}%
\providecommand \@sanitize@url [0]{\catcode `\\12\catcode `\$12\catcode
  `\&12\catcode `\#12\catcode `\^12\catcode `\_12\catcode `\%12\relax}%
\providecommand \@@startlink[1]{}%
\providecommand \@@endlink[0]{}%
\providecommand \url  [0]{\begingroup\@sanitize@url \@url }%
\providecommand \@url [1]{\endgroup\@href {#1}{\urlprefix }}%
\providecommand \urlprefix  [0]{URL }%
\providecommand \Eprint [0]{\href }%
\providecommand \doibase [0]{http://dx.doi.org/}%
\providecommand \selectlanguage [0]{\@gobble}%
\providecommand \bibinfo  [0]{\@secondoftwo}%
\providecommand \bibfield  [0]{\@secondoftwo}%
\providecommand \translation [1]{[#1]}%
\providecommand \BibitemOpen [0]{}%
\providecommand \bibitemStop [0]{}%
\providecommand \bibitemNoStop [0]{.\EOS\space}%
\providecommand \EOS [0]{\spacefactor3000\relax}%
\providecommand \BibitemShut  [1]{\csname bibitem#1\endcsname}%
\let\auto@bib@innerbib\@empty
%</preamble>
\bibitem [{\citenamefont {Maeno}\ \emph {et~al.}(1994)\citenamefont {Maeno},
  \citenamefont {Hashimoto}, \citenamefont {Yoshida}, \citenamefont
  {Nishizaki}, \citenamefont {Fujita}, \citenamefont {Bednorz},\ and\
  \citenamefont {Lichtenberg}}]{maeno1994superconductivity}%
  \BibitemOpen
  \bibfield  {author} {\bibinfo {author} {\bibfnamefont {Y.}~\bibnamefont
  {Maeno}}, \bibinfo {author} {\bibfnamefont {H.}~\bibnamefont {Hashimoto}},
  \bibinfo {author} {\bibfnamefont {K.}~\bibnamefont {Yoshida}}, \bibinfo
  {author} {\bibfnamefont {S.}~\bibnamefont {Nishizaki}}, \bibinfo {author}
  {\bibfnamefont {T.}~\bibnamefont {Fujita}}, \bibinfo {author} {\bibfnamefont
  {J.}~\bibnamefont {Bednorz}}, \ and\ \bibinfo {author} {\bibfnamefont
  {F.}~\bibnamefont {Lichtenberg}},\ }\href@noop {} {\bibfield  {journal}
  {\bibinfo  {journal} {Nature}\ }\textbf {\bibinfo {volume} {372}},\ \bibinfo
  {pages} {532} (\bibinfo {year} {1994})}\BibitemShut {NoStop}%
\bibitem [{\citenamefont {Rice}\ and\ \citenamefont
  {Sigrist}(1995)}]{TMRice_1995}%
  \BibitemOpen
  \bibfield  {author} {\bibinfo {author} {\bibfnamefont {T.~M.}\ \bibnamefont
  {Rice}}\ and\ \bibinfo {author} {\bibfnamefont {M.}~\bibnamefont {Sigrist}},\
  }\href {\doibase 10.1088/0953-8984/7/47/002} {\bibfield  {journal} {\bibinfo
  {journal} {Journal of Physics: Condensed Matter}\ }\textbf {\bibinfo {volume}
  {7}},\ \bibinfo {pages} {L643} (\bibinfo {year} {1995})}\BibitemShut
  {NoStop}%
\bibitem [{\citenamefont {Ishida}\ \emph {et~al.}(1998)\citenamefont {Ishida},
  \citenamefont {Mukuda}, \citenamefont {Kitaoka}, \citenamefont {Asayama},
  \citenamefont {Mao}, \citenamefont {Mori},\ and\ \citenamefont
  {Maeno}}]{ishida1998spin}%
  \BibitemOpen
  \bibfield  {author} {\bibinfo {author} {\bibfnamefont {K.}~\bibnamefont
  {Ishida}}, \bibinfo {author} {\bibfnamefont {H.}~\bibnamefont {Mukuda}},
  \bibinfo {author} {\bibfnamefont {Y.}~\bibnamefont {Kitaoka}}, \bibinfo
  {author} {\bibfnamefont {K.}~\bibnamefont {Asayama}}, \bibinfo {author}
  {\bibfnamefont {Z.}~\bibnamefont {Mao}}, \bibinfo {author} {\bibfnamefont
  {Y.}~\bibnamefont {Mori}}, \ and\ \bibinfo {author} {\bibfnamefont
  {Y.}~\bibnamefont {Maeno}},\ }\href@noop {} {\bibfield  {journal} {\bibinfo
  {journal} {Nature}\ }\textbf {\bibinfo {volume} {396}},\ \bibinfo {pages}
  {658} (\bibinfo {year} {1998})}\BibitemShut {NoStop}%
\bibitem [{\citenamefont {Luke}\ \emph {et~al.}(1998)\citenamefont {Luke},
  \citenamefont {Fudamoto}, \citenamefont {Kojima}, \citenamefont {Larkin},
  \citenamefont {Merrin}, \citenamefont {Nachumi}, \citenamefont {Uemura},
  \citenamefont {Maeno}, \citenamefont {Mao}, \citenamefont {Mori} \emph
  {et~al.}}]{luke1998time}%
  \BibitemOpen
  \bibfield  {author} {\bibinfo {author} {\bibfnamefont {G.~M.}\ \bibnamefont
  {Luke}}, \bibinfo {author} {\bibfnamefont {Y.}~\bibnamefont {Fudamoto}},
  \bibinfo {author} {\bibfnamefont {K.}~\bibnamefont {Kojima}}, \bibinfo
  {author} {\bibfnamefont {M.}~\bibnamefont {Larkin}}, \bibinfo {author}
  {\bibfnamefont {J.}~\bibnamefont {Merrin}}, \bibinfo {author} {\bibfnamefont
  {B.}~\bibnamefont {Nachumi}}, \bibinfo {author} {\bibfnamefont
  {Y.}~\bibnamefont {Uemura}}, \bibinfo {author} {\bibfnamefont
  {Y.}~\bibnamefont {Maeno}}, \bibinfo {author} {\bibfnamefont
  {Z.}~\bibnamefont {Mao}}, \bibinfo {author} {\bibfnamefont {Y.}~\bibnamefont
  {Mori}},  \emph {et~al.},\ }\href@noop {} {\bibfield  {journal} {\bibinfo
  {journal} {Nature}\ }\textbf {\bibinfo {volume} {394}},\ \bibinfo {pages}
  {558} (\bibinfo {year} {1998})}\BibitemShut {NoStop}%
\bibitem [{\citenamefont {Xia}\ \emph {et~al.}(2006)\citenamefont {Xia},
  \citenamefont {Maeno}, \citenamefont {Beyersdorf}, \citenamefont {Fejer},\
  and\ \citenamefont {Kapitulnik}}]{PhysRevLett.97.167002}%
  \BibitemOpen
  \bibfield  {author} {\bibinfo {author} {\bibfnamefont {J.}~\bibnamefont
  {Xia}}, \bibinfo {author} {\bibfnamefont {Y.}~\bibnamefont {Maeno}}, \bibinfo
  {author} {\bibfnamefont {P.~T.}\ \bibnamefont {Beyersdorf}}, \bibinfo
  {author} {\bibfnamefont {M.~M.}\ \bibnamefont {Fejer}}, \ and\ \bibinfo
  {author} {\bibfnamefont {A.}~\bibnamefont {Kapitulnik}},\ }\href {\doibase
  10.1103/PhysRevLett.97.167002} {\bibfield  {journal} {\bibinfo  {journal}
  {Phys. Rev. Lett.}\ }\textbf {\bibinfo {volume} {97}},\ \bibinfo {pages}
  {167002} (\bibinfo {year} {2006})}\BibitemShut {NoStop}%
\bibitem [{\citenamefont {Raghu}\ \emph {et~al.}(2010)\citenamefont {Raghu},
  \citenamefont {Kapitulnik},\ and\ \citenamefont
  {Kivelson}}]{PhysRevLett.105.136401}%
  \BibitemOpen
  \bibfield  {author} {\bibinfo {author} {\bibfnamefont {S.}~\bibnamefont
  {Raghu}}, \bibinfo {author} {\bibfnamefont {A.}~\bibnamefont {Kapitulnik}}, \
  and\ \bibinfo {author} {\bibfnamefont {S.~A.}\ \bibnamefont {Kivelson}},\
  }\href {\doibase 10.1103/PhysRevLett.105.136401} {\bibfield  {journal}
  {\bibinfo  {journal} {Phys. Rev. Lett.}\ }\textbf {\bibinfo {volume} {105}},\
  \bibinfo {pages} {136401} (\bibinfo {year} {2010})}\BibitemShut {NoStop}%
\bibitem [{\citenamefont {Wang}\ \emph {et~al.}(2013)\citenamefont {Wang},
  \citenamefont {Platt}, \citenamefont {Yang}, \citenamefont {Honerkamp},
  \citenamefont {Zhang}, \citenamefont {Hanke}, \citenamefont {Rice},\ and\
  \citenamefont {Thomale}}]{Wang_2013}%
  \BibitemOpen
  \bibfield  {author} {\bibinfo {author} {\bibfnamefont {Q.~H.}\ \bibnamefont
  {Wang}}, \bibinfo {author} {\bibfnamefont {C.}~\bibnamefont {Platt}},
  \bibinfo {author} {\bibfnamefont {Y.}~\bibnamefont {Yang}}, \bibinfo {author}
  {\bibfnamefont {C.}~\bibnamefont {Honerkamp}}, \bibinfo {author}
  {\bibfnamefont {F.~C.}\ \bibnamefont {Zhang}}, \bibinfo {author}
  {\bibfnamefont {W.}~\bibnamefont {Hanke}}, \bibinfo {author} {\bibfnamefont
  {T.~M.}\ \bibnamefont {Rice}}, \ and\ \bibinfo {author} {\bibfnamefont
  {R.}~\bibnamefont {Thomale}},\ }\href {\doibase 10.1209/0295-5075/104/17013}
  {\bibfield  {journal} {\bibinfo  {journal} {Europhysics Letters}\ }\textbf
  {\bibinfo {volume} {104}},\ \bibinfo {pages} {17013} (\bibinfo {year}
  {2013})}\BibitemShut {NoStop}%
\bibitem [{\citenamefont {Scaffidi}\ \emph {et~al.}(2014)\citenamefont
  {Scaffidi}, \citenamefont {Romers},\ and\ \citenamefont
  {Simon}}]{PhysRevB.89.220510}%
  \BibitemOpen
  \bibfield  {author} {\bibinfo {author} {\bibfnamefont {T.}~\bibnamefont
  {Scaffidi}}, \bibinfo {author} {\bibfnamefont {J.~C.}\ \bibnamefont
  {Romers}}, \ and\ \bibinfo {author} {\bibfnamefont {S.~H.}\ \bibnamefont
  {Simon}},\ }\href {\doibase 10.1103/PhysRevB.89.220510} {\bibfield  {journal}
  {\bibinfo  {journal} {Phys. Rev. B}\ }\textbf {\bibinfo {volume} {89}},\
  \bibinfo {pages} {220510} (\bibinfo {year} {2014})}\BibitemShut {NoStop}%
\bibitem [{\citenamefont {Wang}\ \emph {et~al.}(2019)\citenamefont {Wang},
  \citenamefont {Zhang}, \citenamefont {Zhang},\ and\ \citenamefont
  {Wang}}]{PhysRevLett.122.027002}%
  \BibitemOpen
  \bibfield  {author} {\bibinfo {author} {\bibfnamefont {W.-S.}\ \bibnamefont
  {Wang}}, \bibinfo {author} {\bibfnamefont {C.-C.}\ \bibnamefont {Zhang}},
  \bibinfo {author} {\bibfnamefont {F.-C.}\ \bibnamefont {Zhang}}, \ and\
  \bibinfo {author} {\bibfnamefont {Q.-H.}\ \bibnamefont {Wang}},\ }\href
  {\doibase 10.1103/PhysRevLett.122.027002} {\bibfield  {journal} {\bibinfo
  {journal} {Phys. Rev. Lett.}\ }\textbf {\bibinfo {volume} {122}},\ \bibinfo
  {pages} {027002} (\bibinfo {year} {2019})}\BibitemShut {NoStop}%
\bibitem [{\citenamefont {NishiZaki}\ \emph {et~al.}(2000)\citenamefont
  {NishiZaki}, \citenamefont {Maeno},\ and\ \citenamefont
  {Mao}}]{nishizaki2000changes}%
  \BibitemOpen
  \bibfield  {author} {\bibinfo {author} {\bibfnamefont {S.}~\bibnamefont
  {NishiZaki}}, \bibinfo {author} {\bibfnamefont {Y.}~\bibnamefont {Maeno}}, \
  and\ \bibinfo {author} {\bibfnamefont {Z.}~\bibnamefont {Mao}},\ }\href@noop
  {} {\bibfield  {journal} {\bibinfo  {journal} {Journal of the Physical
  Society of Japan}\ }\textbf {\bibinfo {volume} {69}},\ \bibinfo {pages} {572}
  (\bibinfo {year} {2000})}\BibitemShut {NoStop}%
\bibitem [{\citenamefont {Pustogow}\ \emph {et~al.}(2019)\citenamefont
  {Pustogow}, \citenamefont {Luo}, \citenamefont {Chronister}, \citenamefont
  {Su}, \citenamefont {Sokolov}, \citenamefont {Jerzembeck}, \citenamefont
  {Mackenzie}, \citenamefont {Hicks}, \citenamefont {Kikugawa}, \citenamefont
  {Raghu} \emph {et~al.}}]{pustogow2019constraints}%
  \BibitemOpen
  \bibfield  {author} {\bibinfo {author} {\bibfnamefont {A.}~\bibnamefont
  {Pustogow}}, \bibinfo {author} {\bibfnamefont {Y.}~\bibnamefont {Luo}},
  \bibinfo {author} {\bibfnamefont {A.}~\bibnamefont {Chronister}}, \bibinfo
  {author} {\bibfnamefont {Y.-S.}\ \bibnamefont {Su}}, \bibinfo {author}
  {\bibfnamefont {D.}~\bibnamefont {Sokolov}}, \bibinfo {author} {\bibfnamefont
  {F.}~\bibnamefont {Jerzembeck}}, \bibinfo {author} {\bibfnamefont {A.~P.}\
  \bibnamefont {Mackenzie}}, \bibinfo {author} {\bibfnamefont {C.~W.}\
  \bibnamefont {Hicks}}, \bibinfo {author} {\bibfnamefont {N.}~\bibnamefont
  {Kikugawa}}, \bibinfo {author} {\bibfnamefont {S.}~\bibnamefont {Raghu}},
  \emph {et~al.},\ }\href@noop {} {\bibfield  {journal} {\bibinfo  {journal}
  {Nature}\ }\textbf {\bibinfo {volume} {574}},\ \bibinfo {pages} {72}
  (\bibinfo {year} {2019})}\BibitemShut {NoStop}%
\bibitem [{\citenamefont {Li}\ \emph {et~al.}(2021)\citenamefont {Li},
  \citenamefont {Kikugawa}, \citenamefont {Sokolov}, \citenamefont
  {Jerzembeck}, \citenamefont {Gibbs}, \citenamefont {Maeno}, \citenamefont
  {Hicks}, \citenamefont {Schmalian}, \citenamefont {Nicklas},\ and\
  \citenamefont {Mackenzie}}]{li2021high}%
  \BibitemOpen
  \bibfield  {author} {\bibinfo {author} {\bibfnamefont {Y.-S.}\ \bibnamefont
  {Li}}, \bibinfo {author} {\bibfnamefont {N.}~\bibnamefont {Kikugawa}},
  \bibinfo {author} {\bibfnamefont {D.~A.}\ \bibnamefont {Sokolov}}, \bibinfo
  {author} {\bibfnamefont {F.}~\bibnamefont {Jerzembeck}}, \bibinfo {author}
  {\bibfnamefont {A.~S.}\ \bibnamefont {Gibbs}}, \bibinfo {author}
  {\bibfnamefont {Y.}~\bibnamefont {Maeno}}, \bibinfo {author} {\bibfnamefont
  {C.~W.}\ \bibnamefont {Hicks}}, \bibinfo {author} {\bibfnamefont
  {J.}~\bibnamefont {Schmalian}}, \bibinfo {author} {\bibfnamefont
  {M.}~\bibnamefont {Nicklas}}, \ and\ \bibinfo {author} {\bibfnamefont
  {A.~P.}\ \bibnamefont {Mackenzie}},\ }\href@noop {} {\bibfield  {journal}
  {\bibinfo  {journal} {Proceedings of the National Academy of Sciences}\
  }\textbf {\bibinfo {volume} {118}},\ \bibinfo {pages} {e2020492118} (\bibinfo
  {year} {2021})}\BibitemShut {NoStop}%
\bibitem [{\citenamefont {Hicks}\ \emph {et~al.}(2014)\citenamefont {Hicks},
  \citenamefont {Brodsky}, \citenamefont {Yelland}, \citenamefont {Gibbs},
  \citenamefont {Bruin}, \citenamefont {Barber}, \citenamefont {Edkins},
  \citenamefont {Nishimura}, \citenamefont {Yonezawa}, \citenamefont {Maeno}
  \emph {et~al.}}]{hicks2014strong}%
  \BibitemOpen
  \bibfield  {author} {\bibinfo {author} {\bibfnamefont {C.~W.}\ \bibnamefont
  {Hicks}}, \bibinfo {author} {\bibfnamefont {D.~O.}\ \bibnamefont {Brodsky}},
  \bibinfo {author} {\bibfnamefont {E.~A.}\ \bibnamefont {Yelland}}, \bibinfo
  {author} {\bibfnamefont {A.~S.}\ \bibnamefont {Gibbs}}, \bibinfo {author}
  {\bibfnamefont {J.~A.}\ \bibnamefont {Bruin}}, \bibinfo {author}
  {\bibfnamefont {M.~E.}\ \bibnamefont {Barber}}, \bibinfo {author}
  {\bibfnamefont {S.~D.}\ \bibnamefont {Edkins}}, \bibinfo {author}
  {\bibfnamefont {K.}~\bibnamefont {Nishimura}}, \bibinfo {author}
  {\bibfnamefont {S.}~\bibnamefont {Yonezawa}}, \bibinfo {author}
  {\bibfnamefont {Y.}~\bibnamefont {Maeno}},  \emph {et~al.},\ }\href@noop {}
  {\bibfield  {journal} {\bibinfo  {journal} {Science}\ }\textbf {\bibinfo
  {volume} {344}},\ \bibinfo {pages} {283} (\bibinfo {year}
  {2014})}\BibitemShut {NoStop}%
\bibitem [{\citenamefont {Steppke}\ \emph {et~al.}(2017)\citenamefont
  {Steppke}, \citenamefont {Zhao}, \citenamefont {Barber}, \citenamefont
  {Scaffidi}, \citenamefont {Jerzembeck}, \citenamefont {Rosner}, \citenamefont
  {Gibbs}, \citenamefont {Maeno}, \citenamefont {Simon}, \citenamefont
  {Mackenzie} \emph {et~al.}}]{steppke2017strong}%
  \BibitemOpen
  \bibfield  {author} {\bibinfo {author} {\bibfnamefont {A.}~\bibnamefont
  {Steppke}}, \bibinfo {author} {\bibfnamefont {L.}~\bibnamefont {Zhao}},
  \bibinfo {author} {\bibfnamefont {M.~E.}\ \bibnamefont {Barber}}, \bibinfo
  {author} {\bibfnamefont {T.}~\bibnamefont {Scaffidi}}, \bibinfo {author}
  {\bibfnamefont {F.}~\bibnamefont {Jerzembeck}}, \bibinfo {author}
  {\bibfnamefont {H.}~\bibnamefont {Rosner}}, \bibinfo {author} {\bibfnamefont
  {A.~S.}\ \bibnamefont {Gibbs}}, \bibinfo {author} {\bibfnamefont
  {Y.}~\bibnamefont {Maeno}}, \bibinfo {author} {\bibfnamefont {S.~H.}\
  \bibnamefont {Simon}}, \bibinfo {author} {\bibfnamefont {A.~P.}\ \bibnamefont
  {Mackenzie}},  \emph {et~al.},\ }\href@noop {} {\bibfield  {journal}
  {\bibinfo  {journal} {Science}\ }\textbf {\bibinfo {volume} {355}},\ \bibinfo
  {pages} {eaaf9398} (\bibinfo {year} {2017})}\BibitemShut {NoStop}%
\bibitem [{\citenamefont {Oguchi}(1995)}]{PhysRevB.51.1385}%
  \BibitemOpen
  \bibfield  {author} {\bibinfo {author} {\bibfnamefont {T.}~\bibnamefont
  {Oguchi}},\ }\href {\doibase 10.1103/PhysRevB.51.1385} {\bibfield  {journal}
  {\bibinfo  {journal} {Phys. Rev. B}\ }\textbf {\bibinfo {volume} {51}},\
  \bibinfo {pages} {1385} (\bibinfo {year} {1995})}\BibitemShut {NoStop}%
\bibitem [{\citenamefont {Mackenzie}\ \emph {et~al.}(1996)\citenamefont
  {Mackenzie} \emph {et~al.}}]{Mackenzie1996510}%
  \BibitemOpen
  \bibfield  {author} {\bibinfo {author} {\bibfnamefont {A.~P.}\ \bibnamefont
  {Mackenzie}} \emph {et~al.},\ }\href {\doibase DOI:
  10.1016/0921-4534(95)00770-9} {\bibfield  {journal} {\bibinfo  {journal}
  {Physica C: Superconductivity}\ }\textbf {\bibinfo {volume} {263}},\ \bibinfo
  {pages} {510 } (\bibinfo {year} {1996})},\ \bibinfo {note} {proceedings of
  the International Symposium on Frontiers of High - Tc
  Superconductivity}\BibitemShut {NoStop}%
\bibitem [{\citenamefont {Pavarini}\ and\ \citenamefont
  {Mazin}(2006)}]{PhysRevB.74.035115}%
  \BibitemOpen
  \bibfield  {author} {\bibinfo {author} {\bibfnamefont {E.}~\bibnamefont
  {Pavarini}}\ and\ \bibinfo {author} {\bibfnamefont {I.~I.}\ \bibnamefont
  {Mazin}},\ }\href {\doibase 10.1103/PhysRevB.74.035115} {\bibfield  {journal}
  {\bibinfo  {journal} {Phys. Rev. B}\ }\textbf {\bibinfo {volume} {74}},\
  \bibinfo {pages} {035115} (\bibinfo {year} {2006})}\BibitemShut {NoStop}%
\bibitem [{\citenamefont {Pavarini}\ and\ \citenamefont
  {Mazin}(2007)}]{PhysRevB.76.079901}%
  \BibitemOpen
  \bibfield  {author} {\bibinfo {author} {\bibfnamefont {E.}~\bibnamefont
  {Pavarini}}\ and\ \bibinfo {author} {\bibfnamefont {I.~I.}\ \bibnamefont
  {Mazin}},\ }\href {\doibase 10.1103/PhysRevB.76.079901} {\bibfield  {journal}
  {\bibinfo  {journal} {Phys. Rev. B}\ }\textbf {\bibinfo {volume} {76}},\
  \bibinfo {pages} {079901} (\bibinfo {year} {2007})}\BibitemShut {NoStop}%
\bibitem [{\citenamefont {Haverkort}\ \emph {et~al.}(2008)\citenamefont
  {Haverkort} \emph {et~al.}}]{PhysRevLett.101.026406}%
  \BibitemOpen
  \bibfield  {author} {\bibinfo {author} {\bibfnamefont {M.~W.}\ \bibnamefont
  {Haverkort}} \emph {et~al.},\ }\href {\doibase
  10.1103/PhysRevLett.101.026406} {\bibfield  {journal} {\bibinfo  {journal}
  {Phys. Rev. Lett.}\ }\textbf {\bibinfo {volume} {101}},\ \bibinfo {pages}
  {026406} (\bibinfo {year} {2008})}\BibitemShut {NoStop}%
\bibitem [{\citenamefont {Vaugier}\ \emph {et~al.}(2012)\citenamefont
  {Vaugier}, \citenamefont {Jiang},\ and\ \citenamefont
  {Biermann}}]{PhysRevB.86.165105}%
  \BibitemOpen
  \bibfield  {author} {\bibinfo {author} {\bibfnamefont {L.}~\bibnamefont
  {Vaugier}}, \bibinfo {author} {\bibfnamefont {H.}~\bibnamefont {Jiang}}, \
  and\ \bibinfo {author} {\bibfnamefont {S.}~\bibnamefont {Biermann}},\ }\href
  {\doibase 10.1103/PhysRevB.86.165105} {\bibfield  {journal} {\bibinfo
  {journal} {Phys. Rev. B}\ }\textbf {\bibinfo {volume} {86}},\ \bibinfo
  {pages} {165105} (\bibinfo {year} {2012})}\BibitemShut {NoStop}%
\bibitem [{\citenamefont {Bickers}\ and\ \citenamefont
  {White}(1991)}]{PhysRevB.43.8044}%
  \BibitemOpen
  \bibfield  {author} {\bibinfo {author} {\bibfnamefont {N.~E.}\ \bibnamefont
  {Bickers}}\ and\ \bibinfo {author} {\bibfnamefont {S.~R.}\ \bibnamefont
  {White}},\ }\href {\doibase 10.1103/PhysRevB.43.8044} {\bibfield  {journal}
  {\bibinfo  {journal} {Phys. Rev. B}\ }\textbf {\bibinfo {volume} {43}},\
  \bibinfo {pages} {8044} (\bibinfo {year} {1991})}\BibitemShut {NoStop}%
\bibitem [{\citenamefont {Hettler}\ \emph {et~al.}(1998)\citenamefont {Hettler}
  \emph {et~al.}}]{PhysRevB.58.R7475}%
  \BibitemOpen
  \bibfield  {author} {\bibinfo {author} {\bibfnamefont {M.~H.}\ \bibnamefont
  {Hettler}} \emph {et~al.},\ }\href {\doibase 10.1103/PhysRevB.58.R7475}
  {\bibfield  {journal} {\bibinfo  {journal} {Phys. Rev. B}\ }\textbf {\bibinfo
  {volume} {58}},\ \bibinfo {pages} {R7475} (\bibinfo {year}
  {1998})}\BibitemShut {NoStop}%
\bibitem [{\citenamefont {Deisz}\ and\ \citenamefont
  {Slife}(2009)}]{PhysRevB.80.094516}%
  \BibitemOpen
  \bibfield  {author} {\bibinfo {author} {\bibfnamefont {J.~J.}\ \bibnamefont
  {Deisz}}\ and\ \bibinfo {author} {\bibfnamefont {T.}~\bibnamefont {Slife}},\
  }\href {\doibase 10.1103/PhysRevB.80.094516} {\bibfield  {journal} {\bibinfo
  {journal} {Phys. Rev. B}\ }\textbf {\bibinfo {volume} {80}},\ \bibinfo
  {pages} {094516} (\bibinfo {year} {2009})}\BibitemShut {NoStop}%
\bibitem [{\citenamefont {Deisz}\ and\ \citenamefont
  {Kidd}(2017)}]{PhysRevB.95.045122}%
  \BibitemOpen
  \bibfield  {author} {\bibinfo {author} {\bibfnamefont {J.~J.}\ \bibnamefont
  {Deisz}}\ and\ \bibinfo {author} {\bibfnamefont {T.~E.}\ \bibnamefont
  {Kidd}},\ }\href {\doibase 10.1103/PhysRevB.95.045122} {\bibfield  {journal}
  {\bibinfo  {journal} {Phys. Rev. B}\ }\textbf {\bibinfo {volume} {95}},\
  \bibinfo {pages} {045122} (\bibinfo {year} {2017})}\BibitemShut {NoStop}%
\bibitem [{\citenamefont {Freericks}(1994)}]{PhysRevB.50.403}%
  \BibitemOpen
  \bibfield  {author} {\bibinfo {author} {\bibfnamefont {J.~K.}\ \bibnamefont
  {Freericks}},\ }\href {\doibase 10.1103/PhysRevB.50.403} {\bibfield
  {journal} {\bibinfo  {journal} {Phys. Rev. B}\ }\textbf {\bibinfo {volume}
  {50}},\ \bibinfo {pages} {403} (\bibinfo {year} {1994})}\BibitemShut
  {NoStop}%
\bibitem [{\citenamefont {Deisz}\ and\ \citenamefont
  {Kidd}(2011)}]{PhysRevLett.107.277003}%
  \BibitemOpen
  \bibfield  {author} {\bibinfo {author} {\bibfnamefont {J.~J.}\ \bibnamefont
  {Deisz}}\ and\ \bibinfo {author} {\bibfnamefont {T.~E.}\ \bibnamefont
  {Kidd}},\ }\href {\doibase 10.1103/PhysRevLett.107.277003} {\bibfield
  {journal} {\bibinfo  {journal} {Phys. Rev. Lett.}\ }\textbf {\bibinfo
  {volume} {107}},\ \bibinfo {pages} {277003} (\bibinfo {year}
  {2011})}\BibitemShut {NoStop}%
\bibitem [{\citenamefont {Kidd}\ \emph {et~al.}(2005)\citenamefont {Kidd},
  \citenamefont {Valla}, \citenamefont {Fedorov}, \citenamefont {Johnson},
  \citenamefont {Cava},\ and\ \citenamefont {Haas}}]{PhysRevLett.94.107003}%
  \BibitemOpen
  \bibfield  {author} {\bibinfo {author} {\bibfnamefont {T.~E.}\ \bibnamefont
  {Kidd}}, \bibinfo {author} {\bibfnamefont {T.}~\bibnamefont {Valla}},
  \bibinfo {author} {\bibfnamefont {A.~V.}\ \bibnamefont {Fedorov}}, \bibinfo
  {author} {\bibfnamefont {P.~D.}\ \bibnamefont {Johnson}}, \bibinfo {author}
  {\bibfnamefont {R.~J.}\ \bibnamefont {Cava}}, \ and\ \bibinfo {author}
  {\bibfnamefont {M.~K.}\ \bibnamefont {Haas}},\ }\href {\doibase
  10.1103/PhysRevLett.94.107003} {\bibfield  {journal} {\bibinfo  {journal}
  {Phys. Rev. Lett.}\ }\textbf {\bibinfo {volume} {94}},\ \bibinfo {pages}
  {107003} (\bibinfo {year} {2005})}\BibitemShut {NoStop}%
\bibitem [{\citenamefont {Shen}\ \emph {et~al.}(2007)\citenamefont {Shen},
  \citenamefont {Kikugawa}, \citenamefont {Bergemann}, \citenamefont {Balicas},
  \citenamefont {Baumberger}, \citenamefont {Meevasana}, \citenamefont {Ingle},
  \citenamefont {Maeno}, \citenamefont {Shen},\ and\ \citenamefont
  {Mackenzie}}]{PhysRevLett.99.187001}%
  \BibitemOpen
  \bibfield  {author} {\bibinfo {author} {\bibfnamefont {K.~M.}\ \bibnamefont
  {Shen}}, \bibinfo {author} {\bibfnamefont {N.}~\bibnamefont {Kikugawa}},
  \bibinfo {author} {\bibfnamefont {C.}~\bibnamefont {Bergemann}}, \bibinfo
  {author} {\bibfnamefont {L.}~\bibnamefont {Balicas}}, \bibinfo {author}
  {\bibfnamefont {F.}~\bibnamefont {Baumberger}}, \bibinfo {author}
  {\bibfnamefont {W.}~\bibnamefont {Meevasana}}, \bibinfo {author}
  {\bibfnamefont {N.~J.~C.}\ \bibnamefont {Ingle}}, \bibinfo {author}
  {\bibfnamefont {Y.}~\bibnamefont {Maeno}}, \bibinfo {author} {\bibfnamefont
  {Z.-X.}\ \bibnamefont {Shen}}, \ and\ \bibinfo {author} {\bibfnamefont
  {A.~P.}\ \bibnamefont {Mackenzie}},\ }\href {\doibase
  10.1103/PhysRevLett.99.187001} {\bibfield  {journal} {\bibinfo  {journal}
  {Phys. Rev. Lett.}\ }\textbf {\bibinfo {volume} {99}},\ \bibinfo {pages}
  {187001} (\bibinfo {year} {2007})}\BibitemShut {NoStop}%
\bibitem [{\citenamefont {Yamanaka}\ \emph {et~al.}(2022)\citenamefont
  {Yamanaka}, \citenamefont {Okazaki},\ and\ \citenamefont
  {Yaguchi}}]{PhysRevB.105.184507}%
  \BibitemOpen
  \bibfield  {author} {\bibinfo {author} {\bibfnamefont {T.}~\bibnamefont
  {Yamanaka}}, \bibinfo {author} {\bibfnamefont {R.}~\bibnamefont {Okazaki}}, \
  and\ \bibinfo {author} {\bibfnamefont {H.}~\bibnamefont {Yaguchi}},\ }\href
  {\doibase 10.1103/PhysRevB.105.184507} {\bibfield  {journal} {\bibinfo
  {journal} {Phys. Rev. B}\ }\textbf {\bibinfo {volume} {105}},\ \bibinfo
  {pages} {184507} (\bibinfo {year} {2022})}\BibitemShut {NoStop}%
\bibitem [{\citenamefont {Deisz}(2007)}]{PhysRevB.75.064507}%
  \BibitemOpen
  \bibfield  {author} {\bibinfo {author} {\bibfnamefont {J.~J.}\ \bibnamefont
  {Deisz}},\ }\href {\doibase 10.1103/PhysRevB.75.064507} {\bibfield  {journal}
  {\bibinfo  {journal} {Phys. Rev. B}\ }\textbf {\bibinfo {volume} {75}},\
  \bibinfo {pages} {064507} (\bibinfo {year} {2007})}\BibitemShut {NoStop}%
\bibitem [{\citenamefont {Li}\ \emph {et~al.}(2022)\citenamefont {Li},
  \citenamefont {Garst}, \citenamefont {Schmalian}, \citenamefont {Ghosh},
  \citenamefont {Kikugawa}, \citenamefont {Sokolov}, \citenamefont {Hicks},
  \citenamefont {Jerzembeck}, \citenamefont {Ikeda}, \citenamefont {Hu} \emph
  {et~al.}}]{li2022elastocaloric}%
  \BibitemOpen
  \bibfield  {author} {\bibinfo {author} {\bibfnamefont {Y.-S.}\ \bibnamefont
  {Li}}, \bibinfo {author} {\bibfnamefont {M.}~\bibnamefont {Garst}}, \bibinfo
  {author} {\bibfnamefont {J.}~\bibnamefont {Schmalian}}, \bibinfo {author}
  {\bibfnamefont {S.}~\bibnamefont {Ghosh}}, \bibinfo {author} {\bibfnamefont
  {N.}~\bibnamefont {Kikugawa}}, \bibinfo {author} {\bibfnamefont {D.~A.}\
  \bibnamefont {Sokolov}}, \bibinfo {author} {\bibfnamefont {C.~W.}\
  \bibnamefont {Hicks}}, \bibinfo {author} {\bibfnamefont {F.}~\bibnamefont
  {Jerzembeck}}, \bibinfo {author} {\bibfnamefont {M.~S.}\ \bibnamefont
  {Ikeda}}, \bibinfo {author} {\bibfnamefont {Z.}~\bibnamefont {Hu}},  \emph
  {et~al.},\ }\href@noop {} {\bibfield  {journal} {\bibinfo  {journal}
  {Nature}\ }\textbf {\bibinfo {volume} {607}},\ \bibinfo {pages} {276}
  (\bibinfo {year} {2022})}\BibitemShut {NoStop}%
\bibitem [{\citenamefont {Mueller}\ \emph {et~al.}(2023)\citenamefont
  {Mueller}, \citenamefont {Iguchi}, \citenamefont {Watson}, \citenamefont
  {Hicks}, \citenamefont {Maeno},\ and\ \citenamefont
  {Moler}}]{mueller2023constraints}%
  \BibitemOpen
  \bibfield  {author} {\bibinfo {author} {\bibfnamefont {E.}~\bibnamefont
  {Mueller}}, \bibinfo {author} {\bibfnamefont {Y.}~\bibnamefont {Iguchi}},
  \bibinfo {author} {\bibfnamefont {C.}~\bibnamefont {Watson}}, \bibinfo
  {author} {\bibfnamefont {C.}~\bibnamefont {Hicks}}, \bibinfo {author}
  {\bibfnamefont {Y.}~\bibnamefont {Maeno}}, \ and\ \bibinfo {author}
  {\bibfnamefont {K.}~\bibnamefont {Moler}},\ }\href@noop {} {\bibfield
  {journal} {\bibinfo  {journal} {arXiv preprint arXiv:2306.13737}\ } (\bibinfo
  {year} {2023})}\BibitemShut {NoStop}%
\bibitem [{\citenamefont {Grinenko}\ \emph {et~al.}(2021)\citenamefont
  {Grinenko}, \citenamefont {Ghosh}, \citenamefont {Sarkar}, \citenamefont
  {Orain}, \citenamefont {Nikitin}, \citenamefont {Elender}, \citenamefont
  {Das}, \citenamefont {Guguchia}, \citenamefont {Br{\"u}ckner}, \citenamefont
  {Barber} \emph {et~al.}}]{grinenko2021split}%
  \BibitemOpen
  \bibfield  {author} {\bibinfo {author} {\bibfnamefont {V.}~\bibnamefont
  {Grinenko}}, \bibinfo {author} {\bibfnamefont {S.}~\bibnamefont {Ghosh}},
  \bibinfo {author} {\bibfnamefont {R.}~\bibnamefont {Sarkar}}, \bibinfo
  {author} {\bibfnamefont {J.-C.}\ \bibnamefont {Orain}}, \bibinfo {author}
  {\bibfnamefont {A.}~\bibnamefont {Nikitin}}, \bibinfo {author} {\bibfnamefont
  {M.}~\bibnamefont {Elender}}, \bibinfo {author} {\bibfnamefont
  {D.}~\bibnamefont {Das}}, \bibinfo {author} {\bibfnamefont {Z.}~\bibnamefont
  {Guguchia}}, \bibinfo {author} {\bibfnamefont {F.}~\bibnamefont
  {Br{\"u}ckner}}, \bibinfo {author} {\bibfnamefont {M.~E.}\ \bibnamefont
  {Barber}},  \emph {et~al.},\ }\href@noop {} {\bibfield  {journal} {\bibinfo
  {journal} {Nature Physics}\ }\textbf {\bibinfo {volume} {17}},\ \bibinfo
  {pages} {748} (\bibinfo {year} {2021})}\BibitemShut {NoStop}%
\bibitem [{\citenamefont {Yuan}\ \emph {et~al.}(2023)\citenamefont {Yuan},
  \citenamefont {Berg},\ and\ \citenamefont {Kivelson}}]{PhysRevB.108.014502}%
  \BibitemOpen
  \bibfield  {author} {\bibinfo {author} {\bibfnamefont {A.~C.}\ \bibnamefont
  {Yuan}}, \bibinfo {author} {\bibfnamefont {E.}~\bibnamefont {Berg}}, \ and\
  \bibinfo {author} {\bibfnamefont {S.~A.}\ \bibnamefont {Kivelson}},\ }\href
  {\doibase 10.1103/PhysRevB.108.014502} {\bibfield  {journal} {\bibinfo
  {journal} {Phys. Rev. B}\ }\textbf {\bibinfo {volume} {108}},\ \bibinfo
  {pages} {014502} (\bibinfo {year} {2023})}\BibitemShut {NoStop}%
\bibitem [{\citenamefont {Hauck}\ \emph {et~al.}(2023)\citenamefont {Hauck},
  \citenamefont {Beck}, \citenamefont {Kennes}, \citenamefont {Georges},\ and\
  \citenamefont {Gingras}}]{hauck2023competition}%
  \BibitemOpen
  \bibfield  {author} {\bibinfo {author} {\bibfnamefont {J.~B.}\ \bibnamefont
  {Hauck}}, \bibinfo {author} {\bibfnamefont {S.}~\bibnamefont {Beck}},
  \bibinfo {author} {\bibfnamefont {D.~M.}\ \bibnamefont {Kennes}}, \bibinfo
  {author} {\bibfnamefont {A.}~\bibnamefont {Georges}}, \ and\ \bibinfo
  {author} {\bibfnamefont {O.}~\bibnamefont {Gingras}},\ }\href@noop {}
  {\enquote {\bibinfo {title} {Competition between d-wave superconductivity and
  magnetism in uniaxially strained sr2ruo4},}\ } (\bibinfo {year} {2023}),\
  \Eprint {http://arxiv.org/abs/2307.10006} {arXiv:2307.10006
  [cond-mat.supr-con]} \BibitemShut {NoStop}%
\end{thebibliography}
%\end{thebibliography}%
%merlin.mbs apsrev4-1.bst 2010-07-25 4.21a (PWD, AO, DPC) hacked
%Control: key (0)
%Control: author (8) initials jnrlst
%Control: editor formatted (1) identically to author
%Control: production of article title (-1) disabled
%Control: page (0) single
%Control: year (1) truncated
%Control: production of eprint (0) enabled
%

\end{document}